# Phenomenological model of high-$T_c$ superconductivity: a MCS model

A. Mourachkine

Université Libre de Bruxelles, Physique des Solides, CP233, Blvd. du Triomphe, 1050 Brussels, Belgium

We present a phenomenological model of high-$T_c$ superconductivity in hole-doped cuprates: a Magnetic Coupling between Stripes (MCS) model. The MCS model is based upon experimental facts, namely, the presence of (i) stripes, (ii) spin fluctuations, and (iii) two order parameters (for pairing and long-range phase coherence) in hole-doped cuprates. We discuss also the superconductivity in electron-doped cuprates. The MCS model is not applicable to Rhutenate compounds where the superconductivity and magnetism coexist independently.

Doped copper-oxides possess unique physical properties. On the one hand, they *still* have the magnetic properties of their parent compounds which are antiferromagnetic (AF) Mott insulators. On the other hand, the CuO$_2$ planes *already* have metallic properties. Phase separation provides a way to reconcile these properties. Such unique combination leads to the appearance of superconductivity (SC) in cuprates. In addition to the predominant d($x^2$-$y^2$) (hereafter d-wave) order parameter (OP) in hole-doped cuprates, many experiments show the presence of s-wave OP [1,2].

*Stripes.* There is clear evidence for charge stripe [3] formation in Nd-doped LSCO and YBCO [4-6]. The dopant-induced holes choose to segregate in these materials into periodically-spaced stripes which separate AF insulating domains. Recently, it is found that the charge stripes in Nd-doped LSCO are intrinsically metallic [6]. This indicates that stripes, most likely, relate to the occurrence of the SC in cuprates. Otherwise, stripes have to be insulating as a consequence of charge-density-wave instability.

*Spin fluctuations.* The magnetic fluctuations of the Cu spins in AF domains at low frequencies are beginning to achieve the same universality across widely different materials with different values of $T_c$ [5,7]. As for the bulk characteristics, the controlling factor in cuprates seems only to be the hole density in CuO$_2$ planes. Recent neutron studies show that AF spin exchange interactions are most likely responsible for the SC mechanism in cuprates [7,8].

*OPs for pairing and long-range phase coherence.* The SC requires pairing and long-range phase coherence. In conventional SCs, the pairing mechanism and mechanism of establishment of the phase coherence are identical: two electrons in each Cooper pair are attracted by phonons, and the phase coherence among all Cooper pairs is established also by phonons. Both the phenomena occur almost simultaneously at $T_c$. In SC copper-oxides, there is a consensus that, in the underdoped regime, the pairing occurs above $T_c$ when the long-range phase coherence is established, $T_{pair} > T_c$ [9-11]. Since the coherence length in the cuprates is very short (~ 15-20 Å) and the density of the Cooper pairs is very low [10], the Cooper pairs, most likely, use another mechanism, different from the pairing one, in order to establish the long-range phase coherence. This leads to the appearance of two distinct OPs: for pairing and for establishing the long-range phase coherence [9,12]. The phases of the two OPs are locked to each other below $T_c$ [1]. Fig. 1 shows the phase diagram of hole-doped cuprates [9].

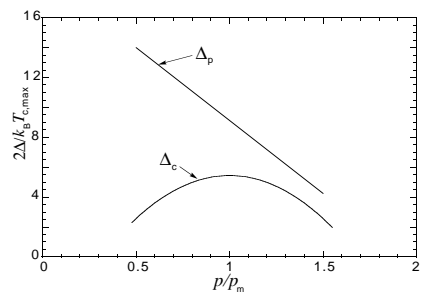

Fig. 1. Phase diagram of hole-doped cuprates at low temperature: $\Delta_c$ is the magnitude of the OP for long-range phase coherence, and $\Delta_p$ is the magnitude of the OP for pairing. The $p_m$ is a hole concentration with the maximum $T_c$ (From Ref. 9).

Tunneling measurements show that the pairing OP in Bi2212 has predominantly SC origin [13]. The comparison of the Josephson $I_cR_n$ products corresponding to the coherent and pairing OPs in Bi2212 shows that the coherent OP has he maximum Josephson strength [14]. The later fact indicate that the coherent OP has the SC origin. Elsewhere [12], we found that, most likely, the coherent OP has the d-wave symmetry while the pairing OP has an anisotropic s-wave symmetry. The coherent OP has most likely the magnetic origin due to spin fluctuations [15].

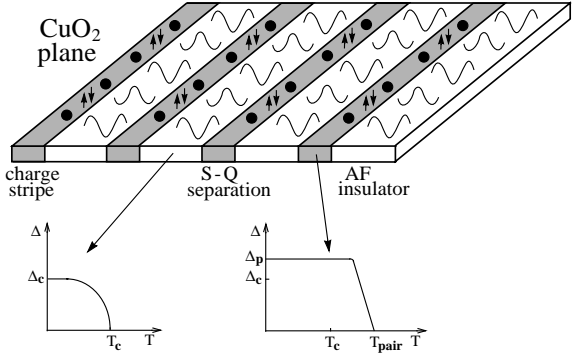

Fig. 2. Static picture of the CuO$_2$ plane in frameworks of the MCS model, shown schematically. Spinons form pairs along charge stripes. The SC mechanism in AF insulating stripes is magnetic due to spin fluctuations. The temperature dependencies are shown schematically.

*Phenomenological model.* Emery with co-workers presented a stripe model of SC in cuprates [16]. In their model, the spinon SC occurs along 1D charge stripes in CuO$_2$ planes at $T_{pair} > T_c$. Spinons form pairs on stripes due to pair hopping between the stripe and its AF environment. The coherent state of the spinon SC is established at $T_c$ by the Josephson coupling between stripes. However, the stripe model can not explain the magnetic origin of the coherent OP. If instead of the Josephson coupling between stripes we introduce a magnetic interstripe coupling due to spin fluctuations, the stripe model may fit the data. Thus, we propose a Magnetic Coupling between Stripes (MCS) model which is entirely based upon experimental facts. In the MCS model, the coherent state of the spinon SC is established due to spin-waves which are excited into local AF domains by dynamically fluctuating stripes. In the MCS model, the SC in cuprates has two mechanisms: along charge stripes and perpendicular to stripes. Charge carriers exhibit different properties in different directions: *fermionic* along charge stripes and *polaronic* perpendicular to stripes. Fig. 2 shows schematically a snapshot of the CuO$_2$ plane in frameworks of the MCS model. The pseudogap in frameworks of the MCS model is similar to that in the stripe model with three characteristic temperatures above $T_c$: $T^*_{stripes} > T^*_{AF} > T_{pair} > T_c$ [16]. $T^*_{stripes}$ and $T^*_{AF}$ correspond respectively to the formation of charge stripes and AF domains between stripes. In frameworks of the MCS model, any impurity (magnetic or nonmagnetic) doped into CuO$_2$ planes affects both the $T_c$ value and density of the states. We found that many experimental data can be explained by using the MCS model [17], including tunneling [17,18] and neutron data [19].

*NCCO.* NCCO and its homologues present a very distinct set of properties with respect to the other cuprates. The analysis of many measurements suggests that the SC in NCCO is mediated by phonons [20,21]. It is possible that, in electron-doped cuprates, stripes carrying spinon SC in order to establish the long-range phase coherence couple to each other due to phonons. In this case, in electron-doped cuprates, a magnetic impurity doped into CuO$_2$ will suppress the SC stronger than a nonmagnetic impurity.

In conclusions, we presented the MCS phenomenological model of the SC in hole-doped cuprates, which is entirely based upon experimental facts. We discussed also the SC in electron-doped cuprates.

This work is supported by PAI 4/10.